# Oxygen Incorporation in the MBE growth of Sc$_x$Ga$_{1-x}$N and Sc$_x$Al$_{1-x}$N

*Joseph Casamento[1*], Prof. Huili Grace Xing[1,2,3], and Prof. Debdeep Jena[1,2,3]*

[1]Department of Materials Science and Engineering, Cornell University, Ithaca, NY 14853, USA
[2]School of Electrical & Computer Engineering, Cornell University, Ithaca, NY 14853, USA
[3]Kavli Institute at Cornell for Nanoscale Science, Cornell University, Ithaca NY 14853, USA

Corresponding author e-mail: jac694@cornell.edu

## Abstract

Secondary-ion Mass Spectrometry (SIMS) is used to determine impurity concentrations of carbon and oxygen in two scandium containing nitride semiconductor multilayer heterostructures: Sc$_x$Ga$_{1-x}$N/GaN and Sc$_x$Al$_{1-x}$N/AlN grown by molecular beam epitaxy (MBE). In the Sc$_x$Ga$_{1-x}$N/GaN heterostructure grown in metal rich conditions on GaN-SiC template substrates with Sc contents up to 28 atomic percent, the oxygen concentration is found to be below $1 \times 10^{19}$/cm$^3$, with an increase directly correlated with the Scandium content. In the Sc$_x$Al$_{1-x}$N-AlN heterostructure grown in nitrogen rich conditions on AlN-Al$_2$O$_3$ template substrates with Sc contents up to 26 atomic percent, the oxygen concentration is found to be between $10^{19}$ to $10^{21}$/cm$^3$, again directly correlated with the Sc content. The increased oxygen and carbon arises during the deposition of scandium alloyed layers.

## 1. Introduction

III-nitride semiconductors such as GaN, AlN, InN and their corresponding alloys have displayed promising optical and electronic device performance. Recently, scandium and other transition metals (e.g. yttrium, niobium, tantalum, manganese, etc.) have gained attention to increase the functionality of nitride semiconductors through piezoelectric, ferroelectric, superconducting, and magnetic behavior.[1] Sc-III nitrides in particular have shown promise for extremely large increases in piezoelectric coefficients and spontaneous polarizations and even ferroelectric behavior.[2-6] These attractive properties have allowed Sc-III nitrides to find use in applications such as bulk acoustic wave (BAW) resonators and microelectromechanical systems (MEMS).[8-10] Reports exist of utilizing magnetron sputtering to deposit thin film $Sc_xAl_{1-x}N$[11-14], and more recently, reports of molecular beam epitaxy (MBE) grown Sc-III nitrides films.[15-22] However, few reports provide information of the oxygen content in the scandium containing films.

Due to the large oxygen affinity of scandium[23], refractory nature of $Sc_2O_3$, and lack of extremely high purity scandium source material that quotes oxygen concentration, integration of scandium into a nitride semiconductor crystal structure without incorporating significant amounts of oxygen and impurities would need special attention. Oxygen incorporation has been readily observed in the binary semiconductor ScN, and has been shown to act as an electron donor and is related to degenerate electron carrier concentrations in ScN films.[24] Moram et al showed that $Sc_2O_3$ can form during sputter-deposition of ScN if the chamber pressure vacuum is raised above $7.5 \times 10^{-7}$ torr, illustrating the need for ultrahigh vacuum conditions to limit oxygen incorporation in ScN.[25] In that study, similar behavior was not seen for Ti or Zr, indicating Sc has a large affinity for oxygen. Here, we present a SIMS studied calibrated by RBS on MBE grown Sc-III nitride semiconductor multilayer heterostructures. The results that indicate the oxygen incorporation into Sc-III nitrides alloys is significant, and correlates with the scandium content. The evidence of the observed densities of oxygen is a first step towards an investigation of its effects on the electronic properties of ScGaN and ScAlN.

## 2. Experimental

$Sc_xGa_{1-x}N$ (x=0.02 to 0.28) and $Sc_xAl_{1-x}N$ (x=0.05 to 0.26) thin films multilayers were grown by MBE in a Veeco® GenXplor system with a base pressure of $1x10^{-10}$ torr. The underlying semi-insulating GaN-SiC and AlN-Al$_2$O$_3$ substrates were cleaned in solvents and HCl and mounted on silicon carrier wafers with indium paste. The substrates were baked at 200˚C in the MBE chamber load lock for 8 hours to remove atmospheric contamination. During growth, the substrates were rotated to ensure film homogeneity. Sc metal source in a W crucible of 99.99 % purity on a rare earth element basis from American Elements was evaporated using a Telemark® electron beam evaporation system in the MBE environment. Flux stability was achieved with an Inficon® electron impact emission spectroscopy (EIES) system by directly measuring the Sc atomic optical emission spectra. Aluminum (99.9999% purity) and gallium (99.99999% purity) were supplied using Knudsen effusion cells. Nitrogen (99.99995%) active species were supplied using a Veeco® RF UNI-Bulb plasma source, with growth pressure of approximately $10^{-5}$ torr. The growth temperature is reported as the heater temperature on the backside of the substrate measured by a thermocouple. In-situ monitoring of film growth was performed using a KSA Instruments reflection high-energy electron diffraction (RHEED) apparatus with a Staib electron gun operating at 15 kV and 1.5 A. The samples were sent to Evans Analytical Group (EAG) for SIMS profiing, which is calibrated by RBS measurements due to the lack of corresponding Scandium standards. Ex-situ atomic force microscopy (AFM) measurements were performed in tapping mode with Veeco Icon and Asylum Research Cypher ES systems.

The $Sc_xGa_{1-x}N$ – GaN heterostructure consisted of four repeat units of 50 nm of GaN followed by 50 nm of $Sc_xGa_{1-x}N$. Nucleation proceeded directly on a semi-insulating (SI) GaN-SiC template substrate, without any buffer layer growth. The substrate GaN surface was first exposed to a gallium flux without $N_2$ at 750°C to desorb surface oxides. The growth temperature was 750°C and the nitrogen plasma condition was 1.95 sccm, 200W. The growth rates were approximately 360 nm/hour. GaN and $Sc_xGa_{1-x}N$ layers were grown metal rich (III/V ratio greater than 1) since Sc has a thermodynamic preference over gallium to bond with nitrogen[26], so the excess gallium remains as a surfactant. This is similar to the typical metal-rich growth conditions that leads to high quality, smooth GaN growth under similar conditions[27]. $Sc_xGa_{1-x}N$ layers were grown by adding the scandium flux to the existing gallium flux. The heterostructure surface was smooth as measured by atomic force microscopy (AFM) with an RMS value of 1.2 nm.

The $Sc_xAl_{1-x}N$ – AlN heterostructure similarly consisted of four repeat units of 65 nm of AlN followed by 75 nm of $Sc_xAl_{1-x}N$. Nucleation proceeded directly on an AlN-$Al_2O_3$ template substrate, with a 360 nm AlN buffer. AlN layers were grown metal rich with excess aluminum consumed[28] before deposition of $Sc_xAl_{1-x}N$. $Sc_xAl_{1-x}N$ layers were grown nitrogen rich, with a III/V ratio of 0.9. Here, III/V ratio is Sc+Al flux ratio to the N* flux ratio, where N* ratio is the flux of nitrogen radical species that contribute to the growth. The N* content at 200W plasma power was calculated from X-Ray Diffraction analysis of the Al content and thickness of $Al_xGa_{1-x}N$-GaN layers (not shown) grown at 200W. The total nitrogen flux also contains $N_2$ molecules that do not contribute to crystal growth, since aluminum does not react with diatomic nitrogen under these temperature and pressure conditions. Nitrogen-rich growth conditions were utilized to promote smooth surfaces and unwanted metallic inclusions, as has been demonstrated in the literature[16,17]. Because Aluminum has thermodynamic preference to bond to nitrogen over scandium[26], excess scandium would be expected to remain on the surface. MBE growth of ScN at growth temperatures above 1000°C (not shown) has indicated that scandium does not desorb at the growth temperatures used in this study. Post-growth AFM showed that the heterostructure surface was rough, with a RMS value of 10 nm. This roughness could originate from the non-ideal growth conditions necessary for the alternating layers of the multilayer: metal-rich for AlN and N-rich for ScAlN. This problem should not appear for a single AlN/ScAlN heterostructure. Indeed, single layer $Sc_xAl_{1-x}N$ films showed smooth surfaces by AFM which is not discussed here; this work focuses on the correlation of impurity incorporation with Sc content, which requires the multilayers.

## 3. Results/Discussion

SIMS results shown in Figure 1 illustrate an order of magnitude increase in oxygen content in scandium-containing layers, followed by a sharp decrease during the GaN layer deposition. In addition, the oxygen concentrations increase as the scandium concentration in the crystal increases and decrease during the GaN layer deposition.

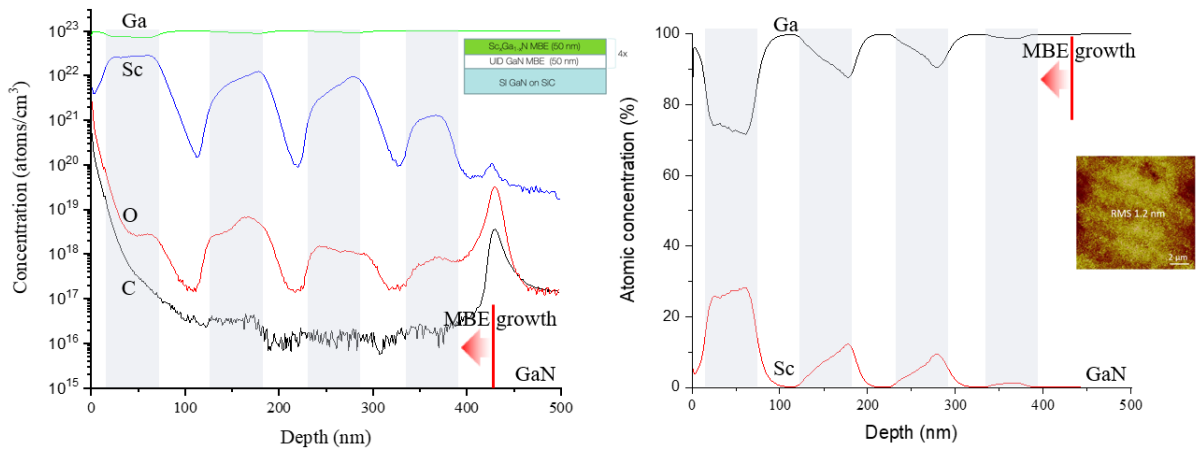

Figure 1: SIMS profile of several elements in a 50/50 nm $Sc_xGa_{1-x}N$/GaN multilayer stack grown by plasma-assisted MBE on a GaN template. Grey areas point to ScGaN layers and white areas point to GaN layers. The left figure shows the concentration of Ga, Sc, O, and C as a function of depth, with the peaks in Sc, O, and C indicating the MBE

nucleation surface on the GaN substrate. The C level is low and does not correlate with the Sc content. The O level is higher, in the $10^{17}$-$8\times10^{18}$/cm$^3$ range, and unlike the C, correlates strongly with the Sc content. The figure on the right shows the Ga and Sc contents in a linear scale, and an AFM image indicating a surface roughness of 1.2 nmThe Sc concentrations are not constant within each layer, indicating either a varying E-Beam Sc flux, or interdiffusion, the origin of which is not clear yet.

These results point to the fact that the increased amount oxygen is coming from the scandium metal and not from residual sources in the MBE chamber. Overall, the oxygen content varies from $3\times10^{17}$/cm$^3$ to $5\times10^{18}$/cm$^3$. This value is higher than prior reports which show that oxygen incorporation in MBE-grown GaN has shown to be reduced to $10^{17}$/cm$^3$ at growth temperatures of 655˚C.[29]

For the corresponding ScAlN/AlN multilayer structure, the SIMS results shown in Figure 2 indicate a large change in oxygen concentration, from $8\times10^{18}$/cm$^3$ near the substrate surface to $3\times10^{21}$/cm$^3$ near the top surface of the film, as the scandium content in the crystal is increased. Similar to the Sc$_x$Ga$_{1-x}$N heterostructure, the oxygen content decreases sharply by over an order of magnitude during the AlN layer deposition. The results again suggest that the increased oxygen concentration in the films is coming from the scandium metal source material.

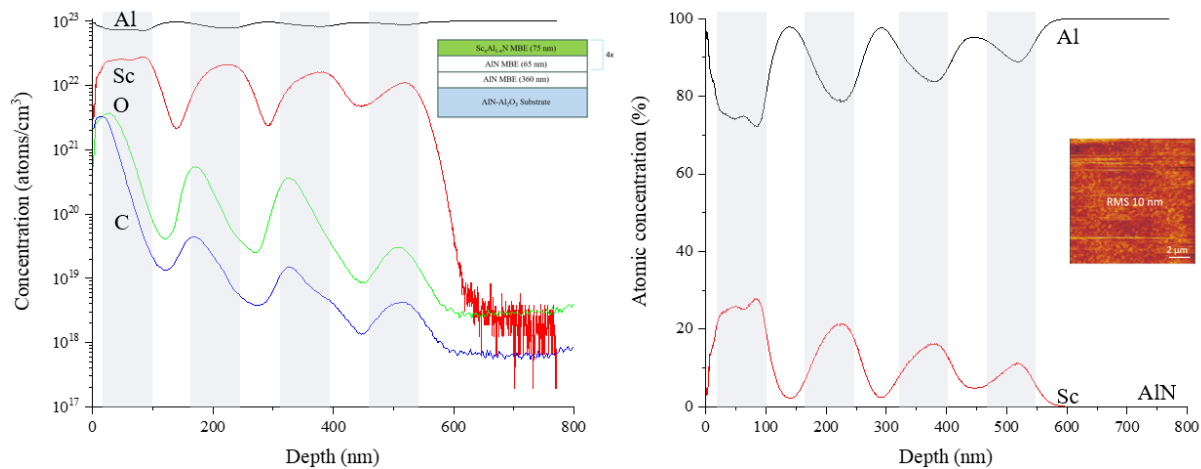

Figure 2: SIMS profile of several elements in a 75/65 nm $Sc_xAl_{1-x}N$/AlN multilayer stack grown by plasma-assisted MBE on an AlN-$Al_2O_3$ substrate. Grey areas point to ScAlN layers and white areas point to AlN layers. The surface morphology of this sample (~10 nm) was rougher than the $Sc_xGa_{1-x}N$ multilayer (~1 nm), as shown in the AFM image on the right. The left figure shows the concentration of Al, Sc, O, and C as a function of depth. The start of the AlN nucleation layer is deeper than the 800 nm plotted. The C level is higher than in $Sc_xGa_{1-x}N$ and is correlated with the Sc content. The O level is quite high, in the $10^{19}$-$3x10^{21}$/cm$^3$ range, and is correlated strongly with the Sc content. The figure on the right shows the Al and Sc contents in a linear scale. The Sc concentrations are not constant within each layer, indicating either a varying E-Beam Sc flux, or interdiffusion, the origin of which is not clear yet.

It is difficult to distinguish between the oxygen that can originate at the substrate surface and oxygen that can originate from the MBE chamber. Accordingly, it is currently unclear what influence the substrate may have on the oxygen concentration in the film. No detectable oxygen ($1x10^{-12}$ torr sensitivity) was measured in the residual gas analyzer (RGA) spectrum of the MBE chamber prior to growth, which gives evidence for oxygen in the films not originating from the MBE chamber. Regarding the scandium metal source, oxygen concentration is not specified by the source vendor. Thus, we cannot determine the oxygen concentration change from the source to the thin film and cannot assess if oxygen is removed during the heating of the scandium metal.

Future work will involve characterization and quantification of the oxygen content in the scandium metal prior to growth.

Regarding the growth, while 750°C has been previously demonstrated to be a reasonable temperature to achieve smooth surfaces and crystalline $Sc_xAl_{1-x}N$, a larger range of growth temperatures need to be conducted to reveal the optimal conditions for reducing oxygen content in the films. Particularly, a focus on higher growth temperatures to reduce oxygen incorporation in $Sc_xAl_{1-x}N$ relative to $Sc_xGa_{1-x}N$ will be conducted. Since $Sc_2O_3$ and $Al_2O_3$ have a larger thermodynamic driving force (e.g. larger negative enthalpy of formation) to form than $Ga_2O_3$, as shown in Figure 3, higher growth temperatures can help facilitate dissociation of metal-oxide bonds

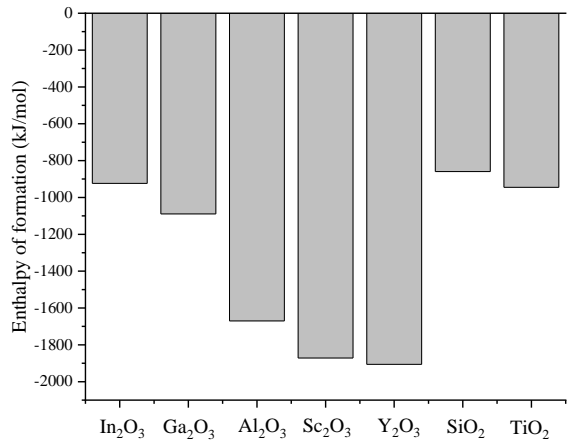

Figure 3: Experimental enthalpy of formation for various binary oxides obtained by calorimetry measurements at 300 K. Data taken from [23].

In addition, $Sc_2O_3$ and $Al_2O_3$ require higher temperatures to reach similar vapor pressures as $Ga_2O_3$. This trend is seen for MBE grown GaN and $Al_xGa_{1-x}N$ layers, where increasing the substrate temperature for 610-655°C shows a large decrease in oxygen content for GaN and an increase from 680°C to 780°C does the same for $Al_xGa_{1-x}N$.[30] Furthermore, it is known that gallium suboxide desorption is prevalent in high-vacuum conditions above approximately 550°C.[31,32] Oxygen concentration in MBE-grown thin films has been shown to decrease as substrate temperature increases. This is in accordance with thermodynamic Ellingham diagrams in that the metal oxide bonds become more unstable as temperature increases and as the partial pressure of oxygen decreases.

For the case of Sc-III nitrides, where oxygen substitutes the nitrogen sublattice, any formation of the non-piezoelectric and non-ferroelectric phase $Sc_2O_3$ is highly undesirable. Oxygen concentrations shown here for $Sc_xAl_{1-x}N$ are on the order of $1 \times 10^{19}$ to $3 \times 10^{21}/cm^3$, which is above the conduction band-edge effective density of states for many semiconductors. Depending on the donor activation energy, this can cause significant unintentional doping of the Sc-containing layers, introducing conduction loss in the piezoelectrics, or electrically shorting and compromising ferroelectricity, which is why a way to reduce the oxygen concentration in such alloys is essential.

## 4. Conclusion

Oxygen incorporation in $Sc_xGa_{1-x}N$-GaN and $Sc_xAl_{1-x}N$-AlN heterostructures measured by SIMS indicate the scandium source introduces significant unintentional oxygen in the film. For Sc contents up to 28% in $Sc_xGa_{1-x}N$ grown metal rich, oxygen concentration is increases up to $5 \times 10^{18}/cm^3$. For Sc contents up to 26% in $Sc_xAl_{1-x}N$ grown nitrogen rich, oxygen concentration increases up to $3 \times 10^{21}/cm^3$. These concentrations and the fact that they are directly correlated with the presence of scandium indicates future studies should find ways to control the oxygen incorporation by the use of higher purity Sc source materials, and an exploration of a wider window of growth conditions that together can keep unintentional oxygen incorporation below densities that compromise the very functionality Sc is being introduced into the nitride crystals for.

**Acknowledgements**

This work was supported in part by NSF DMREF grant 1534303, Cornell's nanoscale facility (grant ECCS-1542081), AFOSR grant FA9550-17-1-0048, NSF DMR-1710298, and the Cornell Center for Materials Research Shared Facilities which are supported through the NSF MRSEC program (DMR-1719875).

**References**

[1]     D. Jena, R. Page, J. Casamento, P. Dang, J. Singhal, Z. Zhang, J. Wright, G. Khalsa, Y. Cho, and H.G Xing, Jpn. J. Appl. Phys. **58**, SC0801 (2019).

[2] A. Teshigahara, K.Y Hashimoto, and M. Akiyama, IEEE International Ultrasonics Symposium (2013).

[3] M. Akiyama, T. Kamohara, K. Kano, A. Teshigahara, Y. Takeuchi, and N. Kawahara, Appl. Phys. Lett. **95**, 162107 (2009).

[4] S. Zhang, D. Holec, W.Y Fu, C.J Humphreys, and M.A Moram, J. Appl. Phys. **114**, 133510 (2013).

[5] M.A Moram and S. Zhang, J. Mater. Chem. A. **2**, 6042 (2014).

[6] S. Fichtner, N. Wolff, F. Lofink, L. Kienle, and B. Wagner, J. Appl. Phys. **125**, 114103 (2019).

[7] M. Schneider, M. DeMiguel-Ramos, A. J. Flewitt, E. Iborra, and U. Schmid, Eurosensors Conference 2017.

[8] V. Pashchenko, S. Mertin, F. Paraspour, J. Li, P. Muralt, and S. Ballandras, EFTF/IFCS Proceedings (2017).

[9] Q. Wang, Y. Lu, S. Mishin, Y. Oshmyansky, and D.A Horsley, JMEMS **26**, 5 (2017).

[10] L. Columbo, A. Kochhar, C. Xu, G. Piazza, S. Mishin, and Y. Oshmyansky, IEEE International Ultrasonics Symposium (2017).

[11] S. Mertin, V. Pashchenko, F. Parsapour, C. Nyffeler, C.S Sandu, B. Heinz, O. Rattunde, G. Christmann, M.A Dubois, and P. Muralt, IEEE International Ultrasonics Symposium (2017).

[12] M. Sinusía Lozano, A. Pérez-Campos, M. Reusch, L Kirste, Th Fuchs, A Žukauskaitė, Z Chen, and G. F. Iriarte, Mater. Res. Express **5**, 036407 (2018).

[13] P. M. Mayrhofer, C. Eisenmenger-Sittner, H. Euchner, A. Bittner, and U. Schmid, Appl. Phys. Lett. **103**, 251903 (2013).

[14] M. Akiyama, T. Tabaru, , K. Nishikubo, A. Teshigahara, and K. Kano, J. Cer. Soc. of Jpn. **118**, 12 (2010).

[15] H. C. L. Tsui, L. E. Goff1, N. P. Barradas, E. Alves, S. Pereira, H. E. Beere, I. Farrer, C. A. Nicoll, D. A. Ritchie, and M. A. Moram, Phys. Status Solidi A **212**,12 (2015).

[16] K. Frei, R. Trejo-Hernández, S. Schütt, L. Kirste, M. Prescher, R. Aidam, S. Müller, P. Waltereit, O. Ambacher, and M. Fiederle, Jpn. J. Appl. Phys. **58**, SC1045 (2019).

[17] M.T Hardy, B.P Downey, N. Nepal, D.F Storm, D.S Katzer, and D.J Meyer, Appl. Phys. Lett.**110**, 162104 (2017).

[18] C. Constantin, H. Al-Brithen, M. B. Haider, D. Ingram, and A. R. Smith, Phys Rev B. 70, **193309** (2004).

[19] S. M. Knoll , S. Zhang , T. B. Joyce , M. J. Kappers , C. J. Humphreys , and M. A. Moram, Phys. Status Solidi A 209, **33** (2012).

[20] H. C. L. Tsui, L. E. Goff1, N. P. Barradas, E. Alves, S. Pereira, H. E. Beere, I. Farrer , C. A. Nicoll, D. A. Ritchie , and M. A. Moram, Phys. Status Solidi A 212, **2837** (2015).

[21] M. A. Moram, Y. Zhang, T. B. Joyce, D. Holec, P. R. Chalker, P. H. Mayrhofer, M. J. Kappers, and C. J. Humphreys, J. Appl. Phys. 106, 113533 (2009).

[22] C. Constantin, M. B. Haider, D. Ingram, N. Sandler, K. Sun, P. Ordejón, and A. R. Smith, J. Appl. Phys. **98**, 123501 (2005).

[23] Mah, A.D., *Heats and free energies of formation of Gallium sesquioxide and Scandium sesquioxide.* US Bureau of Mines, **1962**.

[24] J. Cetnar, A. Reed, S. Badescu, S. Vangala, H. Smith, and D. Look, Appl. Phys. Lett **113**, 192104 (2018).

[25] M.A. Moram, Z.H. Barber, and C.J. Humphreys, Thin Solid Films **516**, 23 (2008).

[26] Landolt-Bornstein, Springer Materials (Springer-Verlag, Berlin/Heidelberg, 2001), Vol. 19A4.

[27] C. Poblenz, P. Waltereit, and J.S. Speck, J. Vac. Sci. & Tec. B. **23**, 1379 (2005).

[28] G. Koblmueller, R. Averbeck, L. Geelhaar, H. Riechert, W. Hösler, and P. Pongratz, J. Appl. Phys. **93**, 9591 (2003).

[29] F. Schuberta, S. Wirthb, F. Zimmermannc, J. Heitmannc, T. Mikolajick, and S. Schmultd, Sci. and Tec. of Adv. Mat. **17**, 1 (2016).

[30] G. Koblmüller, R. M. Chu, A. Raman, U. K. Mishra, and J. S. Speck, J. Appl. Phys **107**, 043527 (2010).

[31] P. Vogt, and O. Bierwagen, Appl. Phys. Lett. **108**, 07201 (2016).

[32] P. Vogt, and O. Bierwagen, Appl. Phys. Lett. **106**, 081910 (2015).